\documentclass[a4paper,12pt]{article}
\usepackage{amsmath,amsthm,amssymb}
\usepackage{subfigure}
\usepackage{verbatim}
\usepackage[english]{babel}
\usepackage{graphicx}
\graphicspath{ {images/} }

\newcommand{\mrm}[1]{\mathrm{#1}}

\newcommand{\ignore}[1]{}

\let\oldsqrt\sqrt
\def\sqrt{\mathpalette\DHLhksqrt}
\def\DHLhksqrt#1#2{%
\setbox0=\hbox{$#1\oldsqrt{#2\,}$}\dimen0=\ht0
\advance\dimen0-0.2\ht0
\setbox2=\hbox{\vrule height\ht0 depth -\dimen0}%
{\box0\lower0.4pt\box2}}

\DeclareFontFamily{OT1}{pzc}{}
\DeclareFontShape{OT1}{pzc}{m}{it}%
              {<-> s * [1.25] pzcmi7t}{}
\DeclareMathAlphabet{\mathpzc}{OT1}{pzc}%
                                 {m}{it}
                                 
\usepackage{xspace}

\newcommand{\e}[1]{\ensuremath{\operatorname{e}^{#1}}}

\begin{document}


\title{\bf Bound on dissipative effects from semileptonic neutral $B$-meson decays}

\author{F. Benatti$^{a,b}$, R. Floreanini$^{b}$, S. Marcantoni$^{a,b}$,\\ P. Pinotti$^{a,b}$,
K. Zimmermann$^{a,b}$\\
\\
\small ${}^a$Dipartimento di Fisica, Universit\`a di Trieste, 
34151 Trieste, Italy\\
\small ${}^b$Istituto Nazionale di Fisica Nucleare, Sezione di Trieste,
34151 Trieste, Italy}

\date{\null}

\maketitle

\begin{abstract}
\noindent
The semileptonic decay asymmetry $\mathcal{A}_{\Delta m}$ is studied within the open quantum systems
approach to the physics of the neutral meson $B^0$-$\overline{B^0}$ system: this extended treatment takes into account
possible non-standard, dissipative effects induced by the presence of an external environment.
A bound on these effects is provided through the analysis of available experimental data 
from the Belle Collaboration.
\end{abstract}

\vskip 1cm

\noindent
{\bf 1. Introduction.} Elementary particle physics is usually formulated using standard quantum mechanics, 
considering all systems as isolated from the external environment. This general framework cannot 
however accommodate all phenomena involving elementary particles, specifically those leading to
decoherence and irreversibility. An extension of the standard treatment is needed in order to
properly describe these effects: it can be physically motivated within the so-called
open system approach to quantum systems \cite{Alicki1}-\cite{Chruscinski}.

Quite in general, an open quantum system can be modelled as a subsystem
$\mathcal{S}$ immersed in an external, large environment $\mathcal{E}$.
Although the time-evolution of the global system $\mathcal{S} + \mathcal{E}$
follows the rules of ordinary quantum mechanics, that of the subsystem alone,
obtained by eliminating the environment degrees
of freedom, is no longer unitary and in general rather involved,
due to the exchange of energy and entropy between $\mathcal{S}$ and $\mathcal{E}$.

The description greatly simplifies when the 
interaction between subsystem and environment can be considered to be weak,
a rather common situation in actual applications.
In this case a mathematically precise modelling of the time evolution of $\mathcal{S}$
alone can be given in terms of linear evolution maps,
the so-called {\it quantum dynamical semigroups} \cite{Alicki1}; they automatically satisfy basic physical
requirements, like forward in time composition law (semigroup property),
entropy increase (irreversibility), complete positivity (it guarantees the
physical consistency of the sub-dynamics in any situation \cite{Benatti-rev}).

This open quantum systems formulation represents a rather universal paradigm
for describing irreversibility and decoherence phenomena:
it has been successfully applied to model noisy and dissipative
effects in quantum optics, atomic and molecular systems,
with applications to quantum information and quantum technology 
\cite{Alicki1}, \cite{Gardiner}-\cite{Petritis}.

In the same vein, it can be applied to study irreversibility and
decoherence phenomena in elementary particle systems,
as these systems can be considered isolated from the environment
only in an idealized setting.
Indeed, more and more investigations point to the intriguing possibility that quantum gravity effects
at Planck's scale or more in general, the microscopic dynamics of extended, fundamental
objects (strings and branes) could effectively act as an external environment, 
inducing non-standard, dissipative effects at low energies.%
\footnote{The original idea that at Planck scale quantum fluctuations of the space geometry 
could destroy the smoothness of the spaceÐtime manifold has been
introduced in \cite{Wheeler} and, since then, further discussed by many authors;
for recent reviews see \cite{Amelino, Plato} and the references therein.}
These new, non-standard phenomena are nevertheless expected to be very small in magnitude and thus difficult
to detect: they are suppressed by at least one inverse power of the Planck mass, 
as a rough dimensional estimate reveals \cite{Benatti0};
in spite of this, they can affect interference phenomena thus becoming within the reach of 
actual detector set-ups. 

In this respect, dedicated neutral meson experiments, both at colliders and
meson factories, appear to be particularly
promising. Indeed, suitable neutral
meson observables turn out to be particularly sensible to the new,
dissipative effects, so that their presence can be experimentally probed 
quite independently from other, non-standard phenomena
\cite{Benatti1}-\cite{Benatti3}.

For completeness,
let us mention that the neutral $B$-meson system has been the focus of investigations involving
the study of other non-standard effects. In particular, efforts \hbox{\cite{Bertlmann1}-\cite{Lenz}} have been
devoted to provide upper bounds on the so-called ``decoherence parameter'', originally introduced in \cite{Eberhard}; 
this constant parametrizes a possible modification of the 
interference term that appears in the expression of quantum mechanical 
probabilities as a consequence of the superposition principle.
As such, this parameter is of kinematical character, not involving directly the $B$-meson dynamics.
In addition, investigations adopting a non-standard time evolutions for the 
neutral $B$-meson system have also appeared in the literature \cite{Bertlmann2,Alok}; 
however, those dynamics appear to be postulated in advance, with little physical insight.
In this respect, our study, based on the theory of open quantum systems,
appears to be rather distinct from those already present
in the literature, and the dissipative effects discussed here have essentially no
bearings with the ones presented in \cite{Bertlmann1}-\cite{Alok}.

More specifically, in the present work, 
we shall discuss in detail how dissipative 
phenomena can affect the dynamics of the meson $B^0$-$\overline{B^0}$ system
by focusing on a specific experimentally accessible observable,
the time dependent flavor asymmetry of semi-leptonic $B^0$ decays $\mathcal{A}_{\Delta m}$, 
that has been the focus of intensive experimental studies at $B$-factories in the recent years
\cite{B-book}.
In particular, the Belle Collaboration has made available
binned, raw data for this asymmetry,
corrected for experimental inefficiencies, with both statistical and systematic uncertainties
\cite{Belle}.
As will shall see, these data, although limited, will be able
to provide interesting bounds on one of the
constants parametrizing the dissipative effects.
In view of the large improvements in the determination of
$\mathcal{A}_{\Delta m}$ and other neutral $B$-meson observables that are expected 
from LHCb and Belle II experiments, 
we are confident that our results will stimulate further interest and additional, more refined studies
on possible dissipative effects in neutral meson physics.

\vskip 1cm

\noindent
{\bf 2. Dissipative dynamics for the neutral meson system.} Within the familiar effective approach, the study 
of the propagation and decay of neutral $B^0$, $\overline{B^0}$ mesons requires
a two-dimensional Hilbert space \cite{Branco,Bigi}. Meson states are represented by $2\times 2$ density matrices,
{\it i.e.} hermitian, positive operators ({\it i.e.} with positive eigenvalues), with constant trace
(at least for unitary evolutions). 
In the 
$|B^0\rangle$, $|\overline{B^0}\rangle$ basis, any neutral meson
state can then be written as
\begin{equation}
\rho=
\begin{bmatrix}
\rho_1&\rho_3\cr
\rho_4&\rho_2
\end{bmatrix}
\ ,
\label{1}
\end{equation}
where $\rho_4\equiv\rho_3^*$, with $*$ signifying complex conjugation.
In the framework of open systems, the time evolution of the state 
$\rho$ is realized in terms of 
linear transformations generated by an equation of the form \cite{Benatti1}-\cite{Benatti3}:
\begin{equation}
{\partial\rho(t)\over\partial t}=-i\frak{H}\, \rho(t)+i\rho(t)\, \frak{H}^\dagger 
+\frak{L}[\rho]\ .
\label{2}
\end{equation}
The first two pieces on the r.h.s. give the standard hamiltonian
contribution, while $\frak{L}$ is a linear map that takes into account the dissipative,
non-standard effects.

The effective hamiltonian $\frak{H}$ includes a non-hermitian part,
\begin{equation}
\frak{H}= M-\frac{i}{2}{\mit\Gamma}\ ,
\label{3}
\end{equation}
with $M$ and $\mit\Gamma$ positive hermitian matrices,
that characterizes the masses and natural widths of the $B$-states.
The entries of $\frak{H}$ can be expressed in terms of its eigenvalues:
$\lambda_H=m_H-{i\over 2}\gamma_H$, $\lambda_L=m_L-{i\over 2}\gamma_L$,
and the complex parameters $p_H$, $q_H$, $p_L$, $q_L$, appearing in the
corresponding (right) eigenstates, 
\begin{equation}
\begin{aligned}
&|B_H\rangle=p_H\, |B^0\rangle + q_H\, |\overline{B^0}\rangle\ ,\qquad\quad
|p_H|^2 + |q_H|^2=1\ ,\cr
&|B_L\rangle=p_L\, |B^0\rangle - q_L\, |\overline{B^0}\rangle\ ,\qquad\quad\ \,
|p_L|^2 + |q_L|^2=1\ .
\end{aligned}
\label{4}
\end{equation}

Even in absence of the additional piece $\frak{L}[\rho]$ in (\ref{2}), probability
is not conserved during the time evolution: $d {\rm Tr}[\rho(t)]/dt\leq0$.
This is due to the presence of a non-hermitian part in the effective 
hamiltonian $\frak{H}$. On the other hand, loss of phase 
coherence shows up only when the piece $\frak{L}[\rho]$ is nonvanishing:
it produces dissipation and transitions from pure states to
statistical mixtures. 

Interestingly, quite independently from the details of the dynamics
in the environment, the form of the map $\frak{L}[\rho]$ is uniquely fixed by
the general physical requirements that the time evolution
it generates should satisfy. First of all, the one parameter (=time) family of linear maps
$\Phi_t$, leading from $\rho(0)$ to $\rho(t)=\Phi_t[\rho(0)]$, 
should transform $B$-meson states into
$B$-meson states, and therefore should map an initial density matrix $\rho(0)$
into a density matrix $\rho(t)$ at time $t$; further, it should
have the property of increasing the (von Neumann)
entropy, $S=-{\rm Tr}[\rho(t)\,\ln\rho(t)]$, of obeying the (semigroup)
composition law, $\Phi_t[\rho(t')]=\rho(t+t')$, for $t,\ t'\geq0$,
of preserving the positivity of $\rho(t)$ for all times.
Actually, in the case of correlated $B$-meson systems, this last property
requires the time evolution $\Phi_t$ to be completely positive,
a much restrictive condition than simple positivity \cite{Alicki1}-\cite{Benatti-rev}.

These requirements allow 
the linear map $\frak{L}[\rho]$ to be fully parametrized in terms of
six real constants,
$a$, $b$, $c$, $\alpha$, $\beta$ and $\gamma$, of dimension of energy,
with $a$, $\alpha$, $\gamma$ non negative; they are not all independent:  
the above mentioned property of complete positivity requires
that they satisfy suitable inequalities \cite{Benatti1,Benatti2}.
These parameters are determined by the time
correlations in the environment; as such, they encode its characteristic
physical properties and can be deduced once the microscopic dynamics
in the environment is given. Here, instead, we take an effective attitude
and treat them as unknown phenomenological parameters, to be determined by the experiment.

A rough evaluation on the magnitude of the dissipative effects induced by
$\frak{L}[\rho]$ can be given on the basis of a general dimensional estimate:
they should be proportional to powers of the typical energy of the system under study,
while suppressed by inverse powers of the characteristic energy
scale of the environment.
Following the general idea that dissipation is induced by
quantum effects at a large, fundamental scale $M_F$ \cite{Benatti0}, 
an upper bound on the magnitude 
of the parameters $a$, $b$, $c$, $\alpha$, $\beta$ and $\gamma$
can be roughly evaluated to be of order
$m^2/M_F$, where $m$ is the neutral $B$-meson mass.
If the non-standard effects have a gravitational origin, the scale 
$M_F$ should coincide with the Planck mass $M_P$, and the previous
upper bound would give: $m^2/M_P\sim 10^{-18}\ {\rm GeV}$.%
\footnote{Dissipative effects can be investigated also in other neutral meson systems, and in particular
in kaon systems \cite{Benatti1,Benatti2,DiDomenico}; however, on the basis of the above dimensional estimates, 
the magnitude of such effects are expected to be further suppressed
by a few orders of magnitude due to the smaller mass of the involved mesons.}

The behaviour in time of physical observables related to the various
neutral meson decay channels can be obtained by solving the evolution equation
(\ref{2}) for an arbitrary initial state $\rho(0)$. 
To this aim, it is convenient to use a vector notation 
and write the matrix $\rho$ as the four-dimensional
vector $|\rho\rangle$, with components
$(\rho_1,\rho_2,\rho_3,\rho_4)$.
Then, the evolution equation (\ref{2})
takes the form of a diffusion equation:
\begin{equation}
{d\over d t}|\rho(t)\rangle=\big[{\cal H}+{\cal L}\big]\, |\rho(t)\rangle\ .
\label{5}
\end{equation}
In the basis in which the hamiltonian contribution is diagonal,
\begin{equation}
{\cal H}=-{\rm diag}\big(\gamma_H,\gamma_L,\Gamma_-,\Gamma_+\big)\ ,
\label{6}
\end{equation}
\begin{equation}
\Gamma_\pm=\Gamma\pm i\Delta m\ ,\qquad \Gamma={\gamma_H+\gamma_L\over2}\ ,
\qquad \Delta m=m_H-m_L\ ,
\label{7}
\end{equation}
the dissipative part $\cal L$ is a $4\times 4$ matrix that can be expressed as follows:
\begin{equation}
{\cal L}=\Gamma\,
\begin{bmatrix}
- D&\phantom{-} D &- C&- C^*\cr
\phantom{-} D&- D &\phantom{-} C&\phantom{-} C^*\cr
- C^*&\phantom{-} C^* & - A &\phantom{-} B(\Delta m/\Gamma)\cr
- C&\phantom{-} C &\phantom{-} B^*(\Delta m/\Gamma)&- A\cr
\end{bmatrix}
\ ,
\label{8}
\end{equation}
in terms of two real, $A$, $D$, and two complex, $B$, $C$, adimensional parameters,
simple linear combinations of the constants $a$, $b$, $c$, $\alpha$, $\beta$ and $\gamma$.%
\footnote{The explicit relations expressing $A$, $B$, $C$, $D$ in terms of
$a$, $b$, $c$, $\alpha$, $\beta$ and $\gamma$ can be found in \cite{Benatti3,Marcantoni}.}

Since, as already observed, the contribution of $\mathcal{L}$ is expected to give
very small effects, one can solve (\ref{5}) in perturbation theory,
$$
|\rho(t)\rangle = e^{{\cal H} t}\ |\rho(0)\rangle
+\int_0^t ds\, e^{{\cal H} (t-s)}\, {\cal L}\,
e^{{\cal H} s}\ |\rho(0)\rangle +\ldots \ ,
$$
retaining only the lowest order terms in the dissipative parameters.
In this way, one can find the time dependence of the components
$\rho_1(t)$, $\rho_2(t)$, $\rho_3(t)$, $\rho_4(t)$
of the vector $|\rho(t)\rangle$, and hence
of the density matrix $\rho(t)$ \cite{Benatti3,Marcantoni,Pinotti}.

These results allow an explicit study of the time evolution of any
physical observable of the $B^0$-$\overline{B^0}$ system.
In the formalism of density matrices, they are
described by suitable hermitian operators $\cal O$. 
Of particular interest are those observables ${\cal O}_f$
that are associated with the decay of a neutral meson into final
states $f$. In the $|B^0\rangle$, $|\overline{B^0}\rangle$ basis,
${\cal O}_f$ is represented by a $2\times2$ matrix,
\begin{equation}
{\cal O}_f=
\begin{bmatrix} 
{\cal O}_1& {\cal O}_3\\
{\cal O}_4& {\cal O}_2
\end{bmatrix}
\ ,
\label{9}
\end{equation}
whose entries can be explicitly written in terms of the two 
independent decay amplitudes
${\cal A}(B^0\rightarrow f)$ and ${\cal A}(\overline{B^0}\rightarrow f)$:
\begin{equation}
\begin{aligned} 
&{\cal O}_1=|{\cal A}(B^0\rightarrow f)|^2\ ,\qquad\qquad
{\cal O}_3=\big[{\cal A}(B^0\rightarrow f)\big]^*\, 
{\cal A}(\overline{B^0}\rightarrow f)\ ,\\
&{\cal O}_2=|{\cal A}(\overline{B^0}\rightarrow f)|^2\ ,\qquad\qquad
{\cal O}_4={\cal A}(B^0\rightarrow f)\, 
\big[{\cal A}(\overline{B^0}\rightarrow f)\big]^*\ .
\end{aligned}
\label{10}
\end{equation}
Its mean value is a physical quantity, directly accessible to the experiment,
whose time evolution is given by 
\begin{equation}
\langle{\cal O}_f\rangle(t)\equiv {\rm Tr}\Big[{\cal O}_f\, \rho(t)\Big]\ .
\label{11}
\end{equation}
This general formula can be used to explicitly compute experimentally relevant
decay rates and asymmetries.

Although, as already mentioned, the basic general idea behind the open
system approach to $B$-meson dynamics is that quantum phenomena at Planck's scale
could produce loss of phase-coherence, it should be stressed that the form
(\ref{5})-(\ref{8}) of the evolution equation is quite independent from the
microscopic mechanism responsible for the dissipative effects;
indeed, as already remarked, quite in general the evolution of any
quantum irreversible process can be effectively modeled in terms
quantum dynamical semigroups. In this respect, the dynamics generated by (\ref{5})
is the best suited for experimental tests: any signal of a non-vanishing
value for some of the parameters appearing in (\ref{8}) would attest
in a model independent way the presence of non-standard, dissipative
effects in $B$-meson physics.%
\footnote{In this respect, we stress again that our attitude is quite different
from that of other recent investigations on ``decoherence'' effects in $B$-meson
physics, {\it e.g.} see \cite{Bertlmann2,Alok,Mavromatos}, where {\it ad hoc}
dynamics are postulated in advance.}

\vskip 1cm

\noindent
{\bf 3. Correlated neutral $B$-mesons semileptonic decays.} The non-standard, 
dissipative effects described by the generalized dynamics (\ref{5})
can be most effectively analyzed in experiments involving
correlated neutral mesons, at meson-factories;
indeed, being quantum mechanical interferometers, these set-ups appear particularly 
suitable for detecting tiny phenomena related to loss of quantum coherence.

In those experiments, 
correlated $B^0$-$\overline{B^0}$ mesons are produced from
the decay of the $\Upsilon(4S)$ spin-1 meson resonance \cite{B-book}.
Since it decays into two spinless bosons, it produces
an antisymmetric spatial state. In the resonance-rest frame, the two neutral 
mesons are produced flying apart with opposite momenta $\vec{p}$; 
in the $|B^0\rangle$, $|\overline{B^0}\rangle$ basis, the resulting state can then be described by:
\begin{equation}
|\psi_A\rangle= \frac{1}{2}\Big(|B^0,-\vec{p}\,\rangle \otimes  
|\overline{B^0},\vec{p}\,\rangle -
|\overline{B^0},-\vec{p}\,\rangle \otimes  |B^0,\vec{p}\,\rangle\Big)\ .
\label{12}
\end{equation}
The corresponding density operator $\rho_A$ is a $4\times 4$ matrix
and it is given by the projector onto the above two-meson state:
$\rho_A=|\psi_A\rangle\langle\psi_A|$.
Its evolution in time can be analyzed using the previously discussed single-meson dynamics: 
once produced in a resonance-decay the two mesons can be considered independent and 
evolve in time each according to the
completely positive map generated by (\ref{5}).
This guarantees that the resulting evolution is completely positive and
of semigroup type, therefore assuring a physically meaningful dynamics
at all times.

As explicitly shown by (\ref{12}),
the two neutral $B$-mesons that come from a decay of the $\Upsilon(4S)$ resonance are 
quantum-mechanically entangled, in a way very similar to that of two spin 1/2
particles coming from a singlet state \cite{Nielsen}. As in that case, correlated
measures on the two particles become physically significant.
Indeed, the typical observables that can be studied at a generic meson-factory
are double decay rates, {\it i.e.} the probabilities 
${\cal P}(f_1,t_1; f_2,t_2)$ that a meson decays
into a final state $f_1$ at proper time $t_1$, while the other meson
decays into the final state $f_2$ at proper time $t_2$.
They can be computed using:
\begin{equation}
{\cal P}(f_1,t_1; f_2,t_2)=
\hbox{Tr}\Big[\big({\cal O}_{f_1}\otimes{\cal O}_{f_2}\big) 
\ \rho_A(t_1,t_2)\Big]\ ,
\label{13}
\end{equation}
where ${\cal O}_{f_1}$, ${\cal O}_{f_2}$ 
represent $2\times2$ hermitian
matrices describing the decay of a single meson into the final
states $f_1$, $f_2$, respectively, while
$\rho_A(t_1,t_2)$ is the the time evolution of the initial density matrix 
$\rho_A=|\psi_A\rangle\langle\psi_A|$
up to $t_1$ for the first meson
and up to $t_2$ for the second.

In practice, in the case of the neutral $B$-mesons,
the short lifetime and rapid $B^0$-$\overline{B^0}$ oscillations 
do not allow a precise enough study of the double time dependence in (\ref{13}).
Indeed, much of the analysis at $B$-meson factories is carried out 
using integrated distributions at fixed time interval $t=t_1-t_2$ \cite{B-book}.
One then focuses on single-time distributions, defined by:
\begin{equation}
{\mit\Gamma}(f_1,f_2;t)\equiv\int_0^\infty dt'\ {\cal P}(f_1,t'+t;f_2,t')\ ,
\qquad t\geq0\ ,
\label{14}
\end{equation}
while for negative $t$, one has:
${\mit\Gamma}(f_1,f_2;-|t|)={\mit\Gamma}(f_2,f_1;|t|)$.
 
It should be stressed that, in presence of an external environment,
the behaviour of ${\mit\Gamma}(f_1,f_2;t)$ is in general quite different
from that obtained in the standard case.
The most striking difference arises when
the final states coincide $f_1=f_2=f$ and $t$ approaches zero. Due to the
antisymmetric character of the initial state $|\psi_A\rangle$ in (\ref{12}), quantum mechanics
predicts a vanishing value for ${\mit\Gamma}(f,f;0)$, while in general
this is not the case for the completely positive dynamics generated
by (\ref{5}). This result reinforces the use of correlated mesons systems for studying 
effects leading to loss of phase coherence and dissipation.

We shall now consider observables connected with $B$-meson decays
into semileptonic final states, $h\ell\nu$, where $h$ stands for any
allowed charged hadronic state, while $\ell$ is a charged lepton and $\nu$
its corresponding neutrino. More specifically, we shall focus on
the asymmetry
\begin{equation}
{\cal A}_{\Delta m}(t)=
\frac{ \big[{\mit\Gamma}(h^+,h^-;t)
+{\mit\Gamma}(h^-,h^+;t)\big] - \big[{\mit\Gamma}(h^+,h^+;t)
+{\mit\Gamma}(h^-,h^-;t)\big]}
{{\mit\Gamma}(h^+,h^+;t) +{\mit\Gamma}(h^-,h^-;t)
+{\mit\Gamma}(h^+,h^-;t) + {\mit\Gamma}(h^-,h^+;t)}\ ,
\label{15}
\end{equation}
which turns out to be particularly sensible to the non-standard,
dissipative effects. As mentioned before, this asymmetry has been
the target of intense experimental studies, as it is one of the
preferred observables for obtaining a precise determination of
the $B$-meson mass difference $\Delta m$ \cite{PDG}.

Let us point out that the evolution equation (\ref{5}) is
manifestly invariant under a phase change
of the basis states $|B^0\rangle$, $|\overline{B^0}\rangle$; 
further, no commitment on possible violations of discrete symmetries ($CP$, $T$  and $CPT$)
or of the so-called $\Delta B=\Delta Q$ rule have so far been made.
Actually, the dependence of the asymmetry ${\cal A}_{\Delta m}$ (and in fact of the other
$B$-meson observables) on the dissipative parameters is very distinctive
and quite different from that of the parameters violating the above
mentioned discrete symmetries. In other terms, the appearance of
environment induced dissipative phenomena can be studied independently
from the other, more familiar symmetry-violating effects.

Nevertheless, for the analysis of the data presented in \cite{Belle}, we shall 
adopt the same attitude usually taken in the experimental determination of 
the mass difference $\Delta m$, assuming no $CPT$- and $CP$-violations in mixing
and the validity of the $\Delta B=\Delta Q$ rule.
In this case, the semileptonic $B$-meson decays become ``flavor tagging'',
since the decay channels $B^0\to h^+\ell^-\overline{\nu}$
and $\overline{B^0}\to h^-\ell^+ \nu$ are forbidden.
In other terms, by looking at the charge of the decay products,
one can reconstruct the flavor of the original decayed meson.

One can than group the single-time integrated distributions appearing
in (\ref{15}) into ``same flavor'',
${\mit\Gamma}^{(SF)}(t)={\mit\Gamma}(h^+,h^+;t) +{\mit\Gamma}(h^-,h^-;t)$,
and ``opposite flavor'',
${\mit\Gamma}^{(OF)}(t)={\mit\Gamma}(h^+,h^-;t) +{\mit\Gamma}(h^-,h^+;t)$,
final state decay rates, so that
\begin{equation}
{\cal A}_{\Delta m}(t)=
\frac{ {\mit\Gamma}^{(OF)}(t) - {\mit\Gamma}^{(SF)}(t) }
{ {\mit\Gamma}^{(OF)}(t) + {\mit\Gamma}^{(SF)}(t) }\ .
\label{16}
\end{equation}
The combinations ${\mit\Gamma}^{(i)}(t)$, $i=SF, OF$, are the quantities directly measured
by the experiment, combined in the asymmetry (\ref{16}) in order to reduce systematic
uncertainties. Note that, due to the decay of the neutral $B$ mesons, the useful events
become necessarily rarer and rarer as the time difference $t$ between the two
meson decays increases. To cope with this drawback, one usually group the data
in suitable time-bins, with increasing width. 
The original data set provided in \cite{Belle} is reproduced in Table 1:
for each bin $n$, the two bin extrema, $t_{in}^{(n)}$-$t_{fin}^{(n)}$, the measured asymmetry 
and the associated combined statistical plus systematic error are explicitly reported.
\begin{table}[htb]
    \centering
    \tabcolsep=.08cm
    \small
    \begin{tabular}[htb]{|r|c|c|}
      \hline
      {bin} 
      &{\hskip .6cm $t_{in}-t_{fin}$ [ps]} 
      &{$\mathcal{A}_{\Delta m}$ and total error}  \\  \hline  
       1&{0.0\,-\,0.5}&{$\phantom{-}1.013 \pm 0.028$}\\ \hline
       2&{0.5\,-\,1.0}&{$\phantom{-}0.916 \pm 0.022$}\\ \hline
       3&{1.0\,-\,2.0}&{$\phantom{-}0.699 \pm 0.038$}\\  \hline
       4&{2.0\,-\,3.0}&{$\phantom{-}0.339 \pm 0.056$}\\  \hline
       5&{3.0\,-\,4.0}&{$-0.136 \pm 0.075$}\\ \hline
       6&{4.0\,-\,5.0}&{$-0.634 \pm 0.084$}\\  \hline
       7&{5.0\,-\,6.0}&{$-0.961 \pm 0.077$}\\  \hline
       8&{6.0\,-\,7.0}&{$-0.974 \pm 0.080$}\\  \hline
       9&{7.0\,-\,9.0}&{$-0.675 \pm 0.109$}\\  \hline
      10&{9.0\,-\,13.0}&{$\phantom{-}0.089 \pm 0.193$}\\  \hline
      11&{13.0\,-\,20.0}&{$\phantom{-}0.243 \pm 0.435$}\\  \hline
      \end{tabular}
    \caption{Time-binned data set for the asymmetry $\mathcal{A}_{\Delta m}$ as given in \cite{Belle}.}
    \label{Table1}
\end{table}

These data can now be used to fit the expression of the asymmetry obtained
from the solution of the dissipative time evolution equation (\ref{5}).
As the data in Table 1 are limited, we shall confine the discussion to a model in which the dissipative
parameter $a$ is vanishing: complete positivity then imposes
$\alpha=\gamma$ and $c=b=\beta=\,0$, so that the dissipative contribution $\cal L$ in (\ref{8})
can be expressed in terms of the single adimensional constant $A=\alpha/\Gamma$,
since $B=D=A$ and $C=\,0$ \cite{Benatti3}. This choice further assures invariance under both $CPT$ and $T$
transformations also in the dissipative part $\cal L$ of the evolution equation (\ref{5}).

Explicit computation then gives \cite{Benatti3,Marcantoni}:
\begin{equation}
\begin{aligned}
  {\mit\Gamma}^{(OF)}(t) - {\mit\Gamma}^{(SF)}(t)
  & = \e{-\Gamma t} \left\{
    \left(1+\frac{A}{1+\frac{\Delta m^2}{\Gamma^2}}\right)  \cos(\Delta m\, t )
    +  \frac{A}{\frac{\Delta m}{\Gamma}
    \left(1+\frac{\Delta m^2}{\Gamma^2}\right)}
    \sin(\Delta m\, t)\right\},\\
  {\mit\Gamma}^{(OF)}(t) + {\mit\Gamma}^{(SF)}(t)
  & = \e{-\Gamma t} \left\{
   (1+A) \cosh\left(\frac{\Delta\Gamma}{2}\, t\right)
    + \frac{2\Gamma A}{\Delta\Gamma}\sinh\left(\frac{\Delta\Gamma}{2}\, t\right)
    \right\},
\end{aligned}
\label{17}
\end{equation}
where, for generality, we have kept non vanishing the decay width difference $\Delta\Gamma=\gamma_H-\gamma_L$.
Neglecting dissipative effects, $A=\,0$, and assuming $\Delta\Gamma=\,0$, from the above
formulas one readily obtain the usual expression for the asymmetry,
${\cal A}_{\Delta m}(t)=\cos(\Delta m\,t)$, used at meson factories 
for the precise determination of the mass difference $\Delta m$.

The expressions in (\ref{17}) can not be directly used for fitting purposes,
as the data in Table~1 do not give the magnitude of the asymmetry as the ratio
of instantaneous decay rates, rather as the ratio of the decay probabilities within the given interval of time
defining each bin; these probabilities are proportional to the number of the relevant decay events
that fall in the corresponding bins.
Then, in order to compare the experimental data given Table~1 with the theoretical predictions, 
one should first integrate
the decay rates ${\mit\Gamma}^{(i)}(t)$, $i=SF, OF$, along the time interval defining each of the eleven data-bins,
\begin{equation}
{\mit\Gamma}^{(i)}\big(t_{in}^{(n)}, t_{fin}^{(n)}\big)=
\int_{t_{in}^{(n)}}^{t_{fin}^{(n)}} dt\, {\mit\Gamma}^{(i)}(t)\ ,\qquad  n=1,2,\ldots, 11\ ,
\label{18}
\end{equation}
and from them reconstruct the corresponding values for the asymmetry:
\begin{equation}
{\cal A}_{\Delta m}^{(n)}=
\frac{{\mit\Gamma}^{(OF)}\big(t_{in}^{(n)}, t_{fin}^{(n)}\big) - {\mit\Gamma}^{(SF)}\big(t_{in}^{(n)}, t_{fin}^{(n)}\big)}
{{\mit\Gamma}^{(OF)}\big(t_{in}^{(n)}, t_{fin}^{(n)}\big) + {\mit\Gamma}^{(SF)}\big(t_{in}^{(n)}, t_{fin}^{(n)}\big)}\ ;
\label{19}
\end{equation}
these are the quantities to be compared with the experimental data in the third column of Table 1.

As a consistency check of the whole procedure, we first analyzed the data set of Table 1 in the hypothesis
of vanishing dissipative effects, $A=\,0$, assuming $\Delta\Gamma=10^{-2}\ {\rm ps}^{-1}$, 
a value consistent both with the predictions
of the Standard Model and the present experimental estimations, leaving
only the mass difference $\Delta m$ as an unknown. Performing a least square fit using the {\tt Root} package,
we obtained $\Delta m=0.504\pm 0.009\ {\rm ps}^{-1}$, with $\chi^2=5.0$ for ten degrees of freedom,
a value perfectly compatible with the one given by the Belle Collaboration \cite{Belle} and the world average \cite{PDG}.

We then performed a two-parameter fit of the data in Table 1, keeping both the dissipative constant $A$
and the mass difference $\Delta m$ as free parameters, while fixing the width $\Gamma$ to its world
average value \cite{PDG}, and assuming again $\Delta\Gamma=10^{-2}\ {\rm ps}^{-1}$. 
The result of the fit gives for $\Delta m$
the same value as before, while for the dissipative parameter one gets:
\begin{equation}
A=(1.5 \pm 8.4)\cdot 10^{-3}\ ,
\label{20}
\end{equation}
with $\chi^2=4.9$ for nine degrees of freedom; the correlation between the two fitted
parameters is also low, and given by: $C\big(\Delta m,\, A\big)=0.14$. In Fig.1 the outcome
of the fit is compared with the experimental data, showing also visually
their agreement.

\begin{figure}[htb]
     \begin{center}
     \includegraphics[scale=0.60]{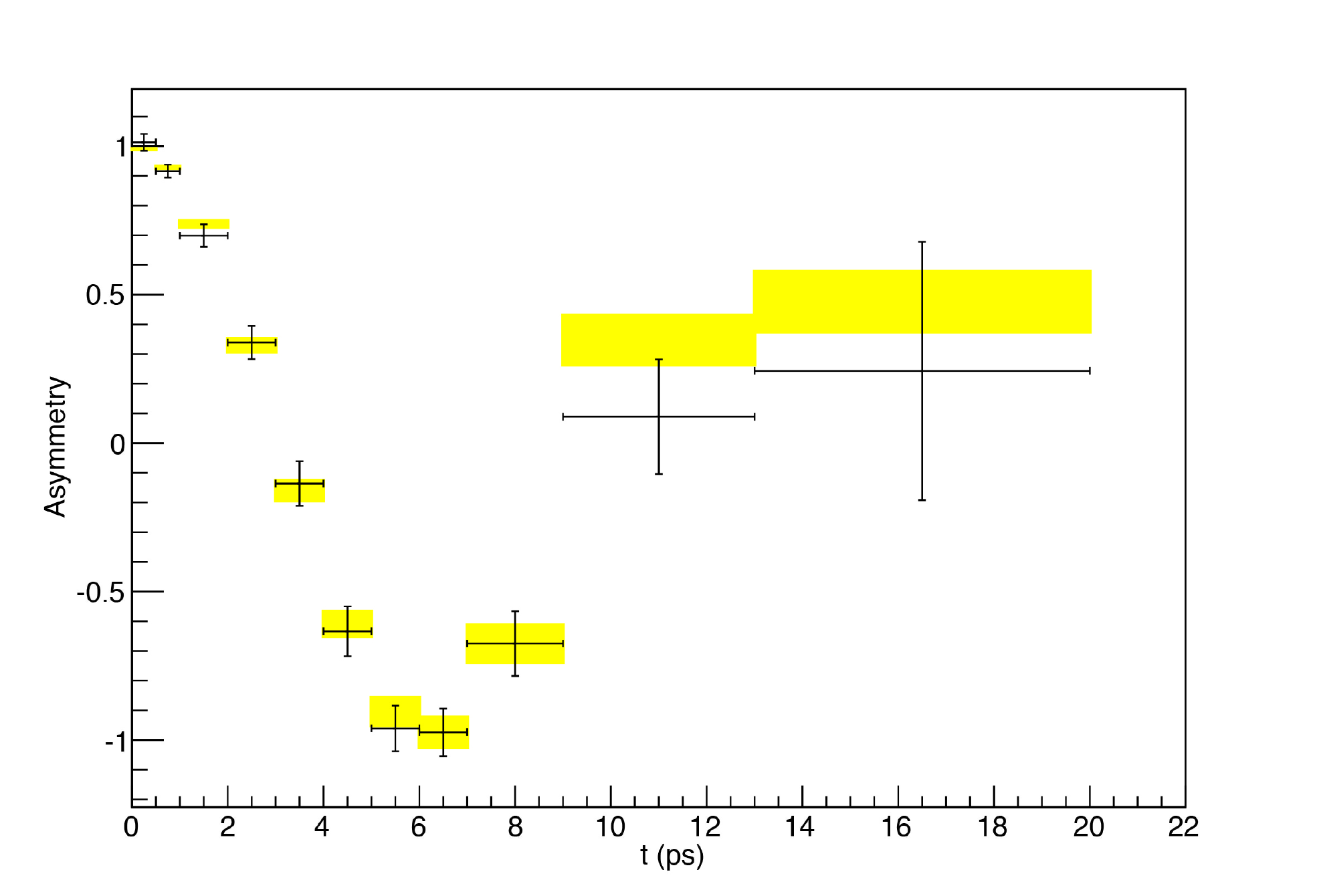}
     \caption{\small Asymmetry ${\cal A}_{\Delta m}$: experimental data (crosses) and the result of the least-squares fit 
(yellow boxes, showing  $\pm 1\sigma$ combined errors on the dissipative parameter $A$ and $\Delta m$).}
     \label{fig: graph}
     \end{center}
 \end{figure}

Although compatible with zero, the obtained value for the parameter $A$ represents one of the best
tests so far available of the presence of dissipative effects in elementary particle physics.
Indeed, going back to the original parameter $\alpha=\Gamma A$ entering the dissipative contribution
to the evolution equation (\ref{2}), one can re-express the result (\ref{20}) as the following upper bound:
$$
\alpha \leq 7.9 \cdot 10^{-15} \, \mrm{GeV}\quad (95\% \, \mrm{CL})\ ,
$$
not very far from the estimate based on quantum gravity effects mentioned before. The future availability
of more accurate data sets and the combined analysis of additional \hbox{$B$-meson} observables will surely
improve this result.

\vskip 1cm

\noindent
{\bf 4. Outlook.} The description of open quantum systems in terms of quantum dynamical
semigroups provides a very general and physically consistent approach to the study
of phenomena leading to irreversibility and dissipation.
When applied to the analysis of the propagation and decay of
correlated neutral $B$-mesons, it gives precise predictions
on the behaviour of relevant physical observables:
the dissipative phenomena can be parametrized through
a set of phenomenological constants,
quite independently from the details of the fundamental,
microscopic dynamics from which they originate.

Various observables
involving the correlated $B^0$-$\overline{B^0}$ system
can be identified as being particularly sensitive to these new phenomena.
In the present investigation, we have focused on the semileptonic decay asymmetry
${\cal A}_{\Delta m}$ and used data from the Belle Collaboration to
give constraints on the dissipative effects; the obtained upper bound
is not too far from estimates based on the hypothesis of a quantum gravity origin 
of these non-standard phenomena.

This is a first step towards a more thorough investigation of environment-induced effects 
in meson systems. Indeed, experimental data related to different observables 
and asymmetries could be readily analyzed with our fully developed theoretical framework.  
These observables will be measured with great accuracy 
at dedicated $B$-meson experiments, both at colliders
(LHCb) and at $B$-factories
(Belle II), so that a much more accurate analysis of the possible presence
of dissipative effects in elementary particle physics can surely be expected in the near future.
In particular, the richer statistics will surely allow for the thorough study of all
six dissipative parameters and the physical implications they underlie.

\end{document}